\documentclass{emulateapj}

\usepackage{amssymb}
\usepackage{natbib}
\usepackage{graphicx}

\newcommand{\req}[1]{Eq.\ (\ref{#1})}
\newcommand{\TF}{T_\mathrm{F}}
 
\begin{document}

\title{Updated Electron-Conduction Opacities: 
The Impact on Low-Mass Stellar Models}

\shorttitle{Updated electron conduction opacities}

\author{S. Cassisi,\altaffilmark{1} A. Y. Potekhin,\altaffilmark{2}
A. Pietrinferni,\altaffilmark{1} M. Catelan,\altaffilmark{3} \& M. Salaris\altaffilmark{4} }

\altaffiltext{1}{INAF -- Osservatorio Astronomico di Teramo, via M.~Maggini s.n., 64100 Teramo, Italy; e-mail:
cassisi, pietrinferni@oa-teramo.inaf.it}
\altaffiltext{2}{Ioffe Physico-Technical Institute,
       Politekhnicheskaya 26, 194021 St.\,Petersburg, Russia;
       e-mail: palex@astro.ioffe.ru}
\altaffiltext{3}{Pontificia Universidad Cat\'olica de Chile,
       Departamento de  Astronom\'\i a y Astrof\'\i sica, 
       Av.\ Vicu\~{n}a Mackenna 4860,
       782-0436 Macul, Santiago, Chile; 
       e-mail: mcatelan@astro.puc.cl}
\altaffiltext{4}{Astrophysics Research Institute, Liverpool John
Moores University, Twelve Quays House, Birkenhead, CH41 1LD, UK; e-mail: ms@astro.livjm.ac.uk}

\slugcomment{Received 2007 January 18; accepted 2007 February 28}

%%%%%%%%%%%%%%%%%%%%%%%%%%%%%%%%%%%%%%%%%%%%%%%%%%%%%%%%%%%%%%%%%%%%%%%%
% Abstract
%%%%%%%%%%%%%%%%%%%%%%%%%%%%%%%%%%%%%%%%%%%%%%%%%%%%%%%%%%%%%%%%%%%%%%%%
\begin{abstract}
We review the theory of electron-conduction opacity, 
a fundamental ingredient in the computation of low-mass stellar models; 
shortcomings and limitations of the existing calculations used in stellar evolution are discussed.  
We then present new determinations of the electron-conduction opacity in stellar conditions for 
an arbitrary chemical composition, that improve over previous works and, most importantly, 
cover the whole parameter space relevant to stellar evolution models 
(i.e., both the regime of partial and high electron degeneracy).    
A detailed comparison with the currently used tabulations   
is also performed. The impact of our new
opacities on the evolution of low-mass stars is assessed 
by computing stellar models along both the H- and He-burning evolutionary phases, as well as  
Main Sequence models of very low-mass stars and white dwarf cooling tracks.

\end{abstract}
\maketitle

%%%%%%%%%%%%%%%%%%%%%%%%%%%%%%%%%%%%%%%%%%%%%%%%%%%%%%%%%%%%%%%%%%%%%%%%
% Introduction
%%%%%%%%%%%%%%%%%%%%%%%%%%%%%%%%%%%%%%%%%%%%%%%%%%%%%%%%%%%%%%%%%%%%%%%%
\section{Introduction}
\label{sect:intro}
One of the main physical inputs required to solve the equations of stellar structure 
is the electron thermal conductivity. When 
the degree of electron degeneracy is significant, electron conduction is the 
dominant energy transport mechanism, and the value of the 
electron-conduction opacity -- proportional 
to the inverse of the electron thermal conductivity -- enters the equation 
of the temperature gradient. This physical condition is verified in the
interiors of brown dwarfs, very low-mass stars with mass $M_\mathrm{tot}<0.15M_\odot$ (the exact value
slightly affected by the metal content;
see \citealt{CB00} for a review),
in the He-core of low-mass stars during their Red Giant Branch (RGB)
evolution (see, e.g., \citealt*{scw02}), in the CO core 
of Asymptotic Giant Branch stars, as well as in white dwarfs 
(inside the core and in a portion of the envelope -- see, e.g., \citealt{PMS02})   
and envelopes of neutron stars. 

Until the mid-1990s, the main sources of electron-conduction opacity 
for stellar model computations were those provided by \citet{HL},
by Itoh and coworkers 
\citep*{FI76,FI79,FI81,I83,I84,mea84,IK93,I93},
and by Yakovlev and coworkers
\citep{YU80,UY80,RY82,Yak87,BY95}.
Each of these sources of opacities has its own shortcomings. 
The \citeauthor{HL} tabulations cover a very limited set of element
mixtures, do not take into account relativistic effects, and contain
significant gaps in the temperature-density diagram (see \S\,\ref{sect:partial} below).
The formulae by \citet{I83,I84,I93} and \citeauthor{IK93},
as well as \citeauthor{YU80}, \citeauthor{RY82}, and \citeauthor{BY95},
are successive improvements -- in their domains of validity --
over the \citeauthor{HL} results for the opacity
contribution due to electron-ion ($ei$) scattering:
they take into account relativistic effects, 
more accurate structure factors for an ion liquid, 
phonon-electron interactions for an ion solid, 
and can be employed 
to compute conductive opacities for arbitrary chemical mixtures.
Simple formulae for the electron-electron ($ee$) scattering contribution
to the conductive opacities have been derived by \citet{UY80} and
\citet*{PCY}.
However, the above-cited results by Itoh's and Yakovlev's groups are applicable   
only when electrons are strongly degenerate ($T\ll\TF$,
where $\TF$ is the Fermi temperature) a condition which is not
strictly satisfied in RGB stars \citep[see, e.g., Fig.~18 in][]{Catelan}.
In addition, they do not take into account important
collective plasma effects near the solid-liquid phase transition
\citep{BKPY}.
These effects have been taken into account by \citet{PBHY},
whose calculations can be also used to compute opacities for arbitrary astrophysical
mixtures at $T\ll\TF$.
For the $ei$ scattering contribution, the latter
calculations
have been extended to partially degenerate plasmas by \citet{VP01},
using a thermal averaging procedure presented by \citet{P99}
(see \S\,\ref{sect:partial} below).

The value of the electron-conduction opacity is 
crucial in particular for the evolution of low-mass RGB stars and the 
fate of their progeny (see, e.g., the extensive 
discussions in \citealt{Catelan} and \citealt*{cdph96}). 
The whole thermal stratification inside the He-core and
the temperature at the base of the surrounding  H-burning shell depend 
strongly on the efficiency of 
electron conduction (for a detailed discussion on this issue see
\citealt{cdph96,scw02}, and references therein). As a consequence,
the mass size of the He core ($M_\mathrm{cHe}$) at the He-burning
ignition (the He-flash at the RGB tip) depends on 
the value of the electron conductivity. 
An accurate determination of the $M_\mathrm{cHe}$ value is
extremely important for several astrophysical problems. First of all,  
the value of $M_\mathrm{cHe}$ regulates the stellar luminosity at the RGB
tip and, in turn, the evolutionary lifetime along the RGB: any increase
of $M_\mathrm{cHe}$ causes an increase in both the surface luminosity at the
RGB tip and the RGB evolutionary lifetime. Given that the mass loss
efficiency along the RGB increases with increasing surface luminosity and decreasing
effective temperature (see, e.g., \citealt{rei75}) the amount
of mass lost during the RGB evolution is strongly influenced by any
variation in the mass of the He core at the He-flash. This 
has a sizeable effect on the morphology of the Horizontal Branch (HB)
(see the review by \citealt{rfp} for a detailed discussion on this
issue). 
In addition, one has to notice that 
the brightness of the RGB tip is a commonly used standard
candle for old, metal-poor stellar systems \citep*{lfm93}. 
Changes in the theoretical predictions of 
$M_\mathrm{cHe}$ have a direct influence on the theoretical calibration of this
distance indicator and, in turn, on the derived
distances \citep{sc98}. Also,  
the luminosity of the core He-burning phase  
in low-mass stars is mainly controlled by the value of 
$M_\mathrm{cHe}$. Given that the HB luminosity
and, in particular, the mean brightness of RR Lyrae stars is 
one of the most
important distance indicators for old stellar systems, 
the theoretical calibration of this standard candle depends on the 
accuracy of the electron-conduction opacities.
 
The value of $M_\mathrm{cHe}$ also affects 
the duration of the core He-burning phase,   
playing therefore a role in the theoretical calibration 
of the so-called $R$-parameter (number ratio of HB stars to RGB 
stars brighter than the HB), which is an indicator of the initial 
He content of old stellar populations \citep{iben68}.
When using  
the Itoh et al.\ conductive opacities, \citet{cd99} found an increase by
$\sim 0.005\,M_\odot$ of $M_\mathrm{cHe}$ at the He flash for a $0.8\,M_\odot$ star with
metallicity $Z=2\times10^{-4}$ compared to models based on
the \citet{HL} opacities. The same numerical experiment performed on a
$1.5\,M_\odot$ star with initial solar composition showed an increase of 
$\sim 0.008\,M_\odot$. A similar
comparison performed with the \citet{PBHY} opacities provides He-core masses 
at the He-ignition between those obtained with 
the Itoh et al.\ and the \citeauthor{HL} ones, but closer 
to the Itoh et al.\ results.

The above-cited papers
by Itoh et al., Yakovlev et al., and Potekhin et al.\ were devoted mostly to 
plasmas at the conditions typical
for neutron star envelopes,
where $ee$ scattering is relatively unimportant.
Therefore the $ee$ scattering mechanism was paid little attention in
those papers.
For instance, \citet{PCY,PBHY} used a fitting formula for 
the $ee$ contribution, valid only for strongly degenerate electrons.
However, as already noted by \citet{HL}
and recently stressed by \citet{Catelan}, the $ee$ scattering
can give a considerable contribution to the opacity in partially
degenerate regions within the He-core of RGB stars. 
In this paper we improve the treatment by \citet{PBHY}
and \citet{P99} for the case of a non-magnetized plasma 
by including the $ee$ scattering in
partially degenerate and nondegenerate matter.
In addition, we take into account a recent improvement of
the treatment of the $ee$ scattering in degenerate matter
by \citet{SY06}. To verify the
impact of these new conductive opacities, we
compute models of low-mass stars and perform a detailed
analysis of the effects of the new conductive opacity evaluations on
both the RGB and HB evolutionary phases. 
We also assess their impact on models of very low-mass stars and white dwarfs.

The paper is organized as follows: in \S~\ref{sect:cond} we briefly discuss the
physical mechanisms governing the electron conduction in the different
physical regimes and emphasize the assumptions made by the various 
existing treatments -- including our new updated results -- with 
the aim of highlighting the limitations of the 
available tabulations of conductive opacities. An analysis of
the impact of our new conductive opacities on stellar models
is presented in \S~\ref{sect:effects}. A summary and final
remarks follow in \S~\ref{sect:concl}.

%%%%%%%%%%%%%%%%%%%%%%%%%%%%%%%%%%%%%%%%%%%%%%%%%%%%%%%%%%%%%%%%%%%%
\section{Stellar Conductive Opacities: the State-of-the-Art}
\label{sect:cond}
Let us recall that
the total opacity of stellar matter \citep[e.g.,][]{rm40,Carson} can be written as 
$\kappa_\mathrm{total}=1/(\kappa_{\mathrm{rad}}^{-1}+\kappa_{\mathrm{c}}^{-1})$,
where $\kappa_{\mathrm{rad}}$ and $\kappa_{\mathrm{c}}$
are the radiative and conductive opacities, respectively.
The latter is related to the thermal conductivity $\varkappa$
by the equation
\begin{equation}
 \kappa_{\mathrm{c}} = \frac{16\sigma T^3}{3\rho\varkappa},
\end{equation}
where $\sigma$ is the Stefan-Boltzmann constant,
$T$ is the temperature, and $\rho$ is the density.

The kinetic method \citep{Ziman} is most practical
for calculating $\varkappa$.
Using the elementary theory
in which 
the effective electron scattering rate
$\nu$ does not depend on the electron
velocity, one can write \citep{Ziman}
\begin{equation}
 \varkappa = a \frac{n_e k^2 T}{m_e \nu},
\label{varkappa}
\end{equation}
where $n_e$ is the electron number density, $m_e$ is the electron mass,
$k$ is the Boltzmann constant,
$a=3/2$ for a nondegenerate electron gas ($T\gg\TF$), and $a=\pi^2/3$
for strongly degenerate electrons ($T\ll\TF$). 
If the electrons are degenerate and relativistic, 
one should replace $m_e$ in \req{varkappa}
by $m_e^* = m_e\gamma_\mathrm{r}$,
where $\gamma_\mathrm{r}=\sqrt{1+x_\mathrm{r}^2}$,
\begin{equation}
 x_\mathrm{r}=p_\mathrm{F}/m_e c = 0.01009 (\rho Z_i/A)^{1/3}
\label{xr}
\end{equation}
is the relativistic density parameter,
$p_\mathrm{F}=\hbar(3\pi^2 n_e)^{1/3}$ is the Fermi momentum, 
$Z_i$ and $A$ are the nuclear charge and mass numbers, 
and the mass density $\rho$ is expressed in g~cm$^{-3}$.
Note that the electron Fermi temperature that accounts for relativistic
effects can be expressed as 
\begin{equation}
   \TF = (m_e^* - m_e) c^2 / k
   = 5.93\times10^9\mbox{~K}
    \left(\gamma_\mathrm{r}-1\right),
\end{equation}
where $k$ is the Boltzmann constant.

In a fully ionized plasma,
$\nu$ is determined by electron-ion and electron-electron
Coulomb collisions. 
Let us assume that 
the effective frequencies of different kinds of collisions simply add up, i.e.,
$\nu=\nu_{ei}+\nu_{ee}$. 
This so-called \emph{Matthiessen rule}
is strictly valid only for
extremely degenerate electrons \citep[see][]{HL}.
However, in practice it gives a good estimate of the 
conductivity: one can show that 
$\nu_{ei}+\nu_{ee}\leq\nu\leq\nu_{ei}+\nu_{ee}+\delta\nu$,
where $\delta\nu\ll\min(\nu_{ei},\nu_{ee})$
\citep[see][]{Ziman}.

\subsection{Nondegenerate Electron Gas}
\label{sect:nondeg}
Let us consider first the case where the electrons
are non-degenerate ($T\gg \TF $)
and non-relativistic ($x_\mathrm{r}\ll1$). 
This case has been studied, e.g.,
by \citet{SpitzerHarm}, \citet{Brag}, and \citet{Spitzer}.
The effective \emph{energy-averaged} $ei$
collision frequency is
\begin{equation}
 \nu_{ei} = \frac43 \sqrt{\frac{2\pi}{m_e}}\,\frac{Z_i^2 e^4}{(kT)^{3/2}}
 \,n_i \Lambda_{ei},
\label{nu_ei_nondeg}
\end{equation}
where $n_i$ is the ion number density, $m_i$ is the ion mass,
and $\Lambda_{ei}$ is the so-called Coulomb logarithm.
In the considered case $\Lambda_{ei}$ is a slowly varying function
of density and temperature.
In general, its value depends on the approximations
used to solve the Boltzmann equation,
but in any case its order-of-magnitude estimate is
given by the elementary theory, where the Coulomb collision integral
is truncated at small and large impact parameters of the electrons.
Then one obtains $\Lambda_{ei}\sim\ln(r_{\mathrm{max}}/r_{\mathrm{min}})$,
where $r_{\mathrm{max}}$ and $r_{\mathrm{min}}$
are the maximum and minimum electron impact parameters.
The parameter $r_{\mathrm{max}}$ can be set equal to
 the Debye screening length,
$r_{\mathrm{max}}^{-2}=4\pi (n_e +Z_i^2 n_i) e^2/kT$.
The second parameter can be estimated as
$r_{\mathrm{min}} = \max(\lambda_T,\,r_{\mathrm{cl}})$,
where $\lambda_T = \sqrt{2\pi\hbar^2/m_e kT}$
is the thermal de Broglie wavelength,
which limits $r_{\mathrm{min}}$ in the high-temperature regime
(where the Born approximation holds),
and $r_{\mathrm{cl}} = Z_ie^2/kT$ is the classical closest-approach
distance of a thermal electron, which limits
$r_{\mathrm{min}}$ in the low-temperature, quasiclassical regime.

A similar effective frequency
\begin{equation}
 \nu_{ee} = \frac83 \sqrt{\frac{\pi}{m_e}}\,\frac{e^4}{(kT)^{3/2}}
 \,n_e \Lambda_{ee}
\label{nu_ee_nondeg}
\end{equation}
characterizes the efficiency of $ee$ collisions.
If $\Lambda_{ee}\sim\Lambda_{ei}$, then $\nu_{ei}/\nu_{ee}\sim Z_i$,
therefore for large $Z_i$ the $ei$ collisions are much more
efficient than the $ee$ collisions.

\subsection{Strongly Degenerate Electrons}
\label{sect:deg}
The thermal conductivity
of strongly degenerate electrons in a fully ionized plasma
is given by \req{varkappa} with $a=\pi^2/3$.
In order to determine the effective collision frequency $\nu$
that enters this equation, we shall 
use the Matthiessen rule, mentioned above.
This will allow us to consider the different types of
electron scatterings separately in the following sections,
and calculate $\nu$ as their sum.

\subsubsection{Electron-Ion Scattering}
Thermal transport coefficients of degenerate electrons
were studied in a number of papers.
In the 1990's, the formulae for conductive opacities mostly used
in astrophysics were those derived by Itoh and coworkers
and by Yakovlev and coworkers (see references in \S\,\ref{sect:intro}).
These formulae are valid for degenerate electrons 
in strongly coupled ion liquids
or solids -- i.e., at
$T\ll \TF $ and $\Gamma>1$, where
\begin{equation}
 \Gamma = \frac{(Z_ie)^2}{a_i kT}
 = \frac{0.2275}{T_6} Z_i^2 \left(\frac{\rho}{A}\right)^{1/3} 
\label{Gamma}
\end{equation}
is the Coulomb coupling parameter\footnote{This parameter
 regulates the state of the Coulomb plasma: the ions constitute a gas at $\Gamma\lesssim1$,
a liquid at $1\lesssim\Gamma<\Gamma_\mathrm{m}$,
and a crystal at $\Gamma>\Gamma_\mathrm{m}$, 
where $\Gamma_\mathrm{m}\approx175$ \citep[see, e.g.,][]{PC00}.}
(here
$a_i = [4\pi n_i/3]^{-1/3}$ is the ion sphere radius,
$T_6 = T/10^6$~K, and $\rho$ is expressed in g~cm$^{-3}$).

It is convenient to present the effective frequency of
$ei$ collisions in the form 
\begin{equation}
 \nu_{ei} = \frac{4\pi Z_i^2 e^4}{p_\mathrm{F}^2 v_\mathrm{F}}\,n_i\Lambda_{ei}
\label{nu_ei_deg}
\end{equation}
(where $v_\mathrm{F} = p_\mathrm{F}/m_e^* = c\,x_\mathrm{r}/\gamma_\mathrm{r}$ is 
the electron Fermi velocity),
and derive a fitting formula for the Coulomb logarithm $\Lambda_{ei}$
as a function of $n_e$, $T$, and $Z_i$.
\citet{YU80} used \req{nu_ei_deg} in the ion liquid regime,
where $\Lambda_{ei}$ is a slow function of order unity.
They employed another form of $\nu_{ei}$ for an ion crystal, 
where collective effects are important -- that is,
where electrons are scattered not by individual ions,
but by their collective excitations, the phonons.

However, \citet{BKPY} showed that in fully ionized dense plasmas
there is no appreciable discontinuity between the conductivities
in the solid and liquid phases. This
is due to the incipient long-range ion structures
in the ion liquid and multiphonon scattering in the solid
near the melting point $\Gamma\sim\Gamma_\mathrm{m}$.
Thus, the conductivity is smooth across the melting.
More recent and more accurate calculations \citep{PBHY}
resulted in new fitting formulae for the $ei$ conductive opacities
in degenerate matter. 
These calculations included 
   the accurate numerical structure factor for a Coulomb liquid 
   calculated by F.~Rogers \& H.~E. DeWitt (unpublished) and fitted 
   by \citet*{YCDW91}, for a Coulomb coupling parameter $\Gamma\geq1$.
\citet{PBHY} have taken account of the modifications of the \emph{effective}
(for electron scattering)
ion structure factor 
around the melting point, discussed by \citeauthor{BKPY},
as well as the quantization of ion motion at
$T$ below the ion plasma temperature $T_\mathrm{p,ion}$
 and ``freezing-out'' of the umklapp processes at $T\ll T_\mathrm{p,ion}$ \citep{RY82}.
A correct limiting form of the ion structure factor
at $\Gamma\ll1$ was also ensured.
Hence the results of \citet{PBHY} are probably valid
at any $\Gamma$. On the other hand, it should be noted that rigorous calculations 
for $\Gamma \la 1$ have never been carried out, and would certainly prove 
of interest in placing this ``smooth interpolation argument'' on a more solid 
physical basis.

The absence of a big discontinuity at $\Gamma=\Gamma_\mathrm{m}$
 allowed \citet{PBHY} to construct a single fitting formula
for $\Lambda_{ei}$ using \req{nu_ei_deg} at all temperatures.
We shall follow this approach hereafter.
It should be noted, however, that 
$\Lambda_{ei}$ in \req{nu_ei_deg} is a slowly varying function
only at $\Gamma\lesssim\Gamma_\mathrm{m}$,
but it rapidly decreases in the solid regime at $\Gamma\gg\Gamma_\mathrm{m}$
(see \citealt{PBHY} for details).

\subsubsection{Electron-Electron Scattering}
\label{sect:ee}
The expression of $\nu_{ee}$ for the relativistic degenerate electrons
at $T\ll T_\mathrm{p}$ was obtained by \citet{FI76}.
Here $T_\mathrm{p}=\hbar\omega_\mathrm{p}/k$ is the electron plasma temperature,
determined by the electron plasma frequency
$\omega_\mathrm{p} = \sqrt{4\pi e^2 n_e / m_e^*}$.
\citet{UY80} extended the results of \citet{FI76}
to higher temperatures, where $T\ll \TF $, but not necessarily
$T\ll T_\mathrm{p}$. 
\citet{PCY,PBHY} 
calculated $ee$ conductive opacities
according to the theory of \citet{UY80}
and presented a fitting formula for their results.

Recently, \citet{SY06} have reconsidered the problem taking 
into account the Landau damping of transverse plasmons.
This effect 
is due to the difference of the components
of the polarizability tensor, responsible for screening
the charge-charge and current-current interactions:
the transverse current-current interactions undergo 
``dynamical screening.''
It was neglected by \citet{UY80} 
but later studied by \citet{HeiselbergPethick}
in the context of the transport coefficients of the quark-gluon plasma.
\citeauthor{SY06} showed that the Landau damping of transverse plasmons
strongly increases $\nu_{ee}$ in the domain
of $x_\mathrm{r} \gtrsim 1$ and $T\ll T_\mathrm{p}$.
Their result can be written as
\begin{equation}
 \nu_{ee} = \frac{m_e c^2}{\hbar}\frac{6\alpha_{\mathrm{f}}^{3/2}}{\pi^{5/2}}\,
 x_\mathrm{r} y \sqrt{\beta_\mathrm{r}}\,I(\beta_\mathrm{r},y),
% =1.66\times10^{17}x_\mathrm{r} y \sqrt{\beta_\mathrm{r}}\,I(\beta_\mathrm{r},y) {\rm~s}^{-1} ,
\label{nu_ee_deg}
\end{equation}
where $\alpha_{\mathrm{f}}$ is the fine-structure constant,
$\beta_\mathrm{r} = v_\mathrm{F}/c = x_\mathrm{r}/\gamma_\mathrm{r}$, 
$y=\sqrt{3}\, T_\mathrm{p}/T = (571.6/T_6) \sqrt{\beta_\mathrm{r}}\,x_\mathrm{r}$, and 
\begin{eqnarray}&&\hspace*{-2em}
 I(\beta,y) = \frac{1}{\beta} \left(\frac{10}{63}-\frac{8/315}{1+0.0435y}\right)
 \nonumber\\&&\times
 \ln\left(1+\frac{128.56}{37.1y+10.83y^2+y^3}\right)
\nonumber\\&&\hspace*{-2em}
 + \beta^3\left(\frac{2.404}{B}+\frac{C-2.404/B}{1+0.1\beta y}\right)
 \nonumber\\&&\times
 \ln\left[1+\frac{B}{A\beta y +(\beta y)^2} \right]
\nonumber\\&&\hspace*{-2em}
 + \frac{\beta}{1+D}\left(C+\frac{18.52\beta^2 D}{B}\right)
 \nonumber\\&&\hspace*{-1em}\times
 \ln\left[1+\frac{B}{Ay+10.83(\beta y)^2 + (\beta y)^{8/3}}\right],
\hspace*{2em}
\label{ee-fit}
\end{eqnarray}
where $A=12.2+25.2\,\beta^3$, $B=A\exp[(0.123636+0.016234\,\beta^2)/C]$,
$C=8/105+0.05714\,\beta^4$, and $D=0.1558\,y^{1-0.75\,\beta}$.

These equations are valid only for $T\ll\TF$;
their extension to the case $T\gtrsim \TF$ will be suggested in
\S\,\ref{sect:partial}.

It is easy to check that Eqs.~(\ref{nu_ee_deg}) and (\ref{ee-fit})
reproduce the low-density ($x_\mathrm{r}\ll1$) asymptote
of the previous fit to $\nu_{ee}$ presented by
\citet{PCY,PBHY}. However, the deviation from the previous results is already
large at $x_\mathrm{r}\sim1$ and can reach many orders of
magnitude at $x_\mathrm{r}\gg1$.

\subsubsection{Impurities and Mixtures}
Real stellar plasmas are often mixtures of different chemical elements. 
In this case \req{xr} for the relativity parameter $x_\mathrm{r}$ has to
be modified by replacing $Z_i/A$ by the unweighted
mean value of the charge-to-mass
ratio, $\langle Z_i/A \rangle$, averaged over all species.
The effective
collision frequency $\nu_{ei}$ should also be modified. 
The required modification can be different, depending on the 
state of the plasma and on the amount of impurities.
For example, \citet{FI76},
\citet{YU80}, and \citet{IK93} considered electron scattering by charged
impurities in a Coulomb crystal. 
If the fraction of impurities is small and they are randomly
distributed, electron-impurity scattering
can be treated as scattering by charge fluctuations, controlled by
the impurity parameter
$
Q= \sum_j Y_j (Z_j-\langle Z\rangle)^2 ,
$
where $Y_j=n_j/\sum_j n_j$ is the number fraction of impurities of the $j$th kind,
$Z_j$ is their charge number, and $\langle Z\rangle$
is the mean charge number, which in the considered case 
is close to the charge number
of the main ion species that forms the crystal lattice.
Then, using the Matthiessen rule, one can obtain $\nu_{ei}$
as a sum of the terms corresponding to the $ei$ (electron-phonon)
scattering in a homogeneous
lattice and to the electron scattering by charge fluctuations,
which is given by \req{nu_ei_deg} with $Z_i$ replaced by $Z_\mathrm{imp}=\sqrt{Q}$.
This approach has been adopted by \citet{YU80}, \citet{PY96}, 
and \citet*{GYP}, and implemented in the online database
referenced below.

An alternative approach is relevant when there is no dominant ion species 
which forms a crystal (e.g., in a liquid, a gas, or a glassy alloy).
In this case, one can use \req{nu_ei_deg} 
with $Z_i^2 n_i \Lambda_{ei}$ replaced by 
$\sum_j Z_j^2 n_j \Lambda_{ei}^j$,
where summation is over all ion species $j$,
and the Coulomb logarithm $\Lambda_{ei}^j$ depends generally on $j$.
An approximation to $\Lambda_{ei}^j$ based on the plasma 
``additivity rule'' has been suggested by \citet{PBHY}.
A much simpler yet reasonable approximation
to the conductive opacity can be obtained by using
an effective charge number equal to $\sqrt{\langle Z^2\rangle}\equiv(\sum_j Z_j^2 n_j)^{1/2}$.

A brief discussion and comparison of the two approaches
has been given by \citet*{BBC}. 

In the present paper, we consider pure He composition for the core of RGB stellar models.
This is justified by low metallicities 
in most of these models, which make the exact conductivity
of a mixture close to the pure He conductivity.
For instance, an admixture of 0.1\% (by number) of N or C to the He liquid
raises the conductive opacity by $\approx1$\%, which can be safely neglected.

\subsection{Partially Degenerate Electrons}
\label{sect:partial}
The thermal conductivity of partially degenerate electrons
is more difficult to calculate. For the fully ionized Coulomb plasmas
this problem was studied by \citet{Lampe68},
who calculated the transport coefficients of a weakly coupled plasma 
of nondegenerate ions and partially degenerate nonrelativistic electrons
by means of a Chapman-Enskog solution of the quantum Lenard-Balescu 
kinetic equation, including both $ee$ and $ei$ collisions.
The employed method is more fundamental and accurate
than the simple estimates presented in Sect.~\ref{sect:nondeg}.
However, the ion correlations have not been taken into account,
i.e., $\Gamma\ll1$ was assumed.
\citet{HL} combined these calculations with earlier 
results of \citet{Hubbard66}, who considered the transport
coefficients determined by the $ei$ collisions
in a nonrelativistic degenerate electron gas, taking into account 
ion-ion correlations. 
\citeauthor{HL} provided conductive opacities in tabular form
for various chemical compositions.

In the domains of degenerate electrons and strongly correlated
ions, more accurate results have been obtained 
after the work of \citet{HL}, as discussed in Sect.~\ref{sect:deg}.
Although the \citeauthor{HL} results are sufficiently
accurate at $T\gg \TF $, their tabular form of presentation
does not allow their extension to 
other chemical compositions or plasma parameters
beyond the tabulated range.

\citet{P99} proposed a procedure of averaging 
the expressions for $ei$ transport coefficients,
obtained in the limit $T/\TF\to0$,
over the thermal distribution of the electron energies.
Although this procedure was
originally devised for taking account of thermal
broadening of oscillations of the transport coefficients
in quantizing magnetic fields at $T\ll \TF $, 
it was shown to reproduce also the correct
non-magnetic limit of $\varkappa$ at $T\gg \TF $ \citep{VP01}.
According to this recipe,
\begin{equation}
 \varkappa = k^2 T (\sigma_2-\sigma_1^2/\sigma_0),
\label{thav-kappa}
\end{equation}
where
\begin{equation}
 \sigma_n = \int \frac{\chi^n}{\nu_{ei}(\epsilon)}
 \frac{\mathcal{N}(\epsilon)}{m_e^*(\epsilon)}
 \frac{e^\chi}{(e^\chi+1)^2}\,{\mathrm{d}}\chi,
\label{thav}
\end{equation}
$\epsilon$ is the electron energy,
$\chi=(\epsilon - \mu)/kT$, 
$\mu$ is the electron chemical potential,
$\mathcal{N}(\epsilon)$ is the number density
of electron states with energies smaller than $\epsilon$
(i.e., the electron number density that would correspond to
the Fermi energy $\epsilon$ at $T=0$),
and $\nu_{ei}(\epsilon)$ is the effective collision
frequency of electrons with energy $\epsilon$.
In a nonquantizing (or zero) magnetic field,
$\mathcal{N}(\epsilon)=p^3/(3\pi^2\hbar^3)$
and $m_e^*(\epsilon)=\sqrt{m_e^2+(p/c)^2}$,
where $p$ is the electron momentum.

We adopt this approach to the $ei$ opacity
calculations at arbitrary $T/\TF $.

%%% FIG 1

\begin{figure} \epsscale{.9}
\plotone{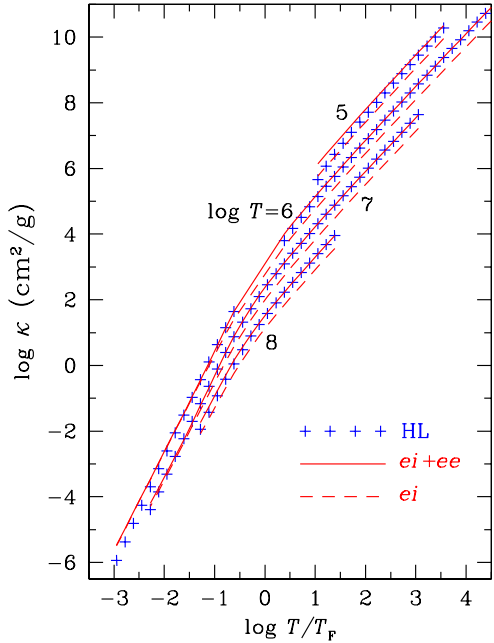}
\caption{Conductive opacity 
($\kappa \equiv \kappa_\mathrm{c}=\kappa_{ei}+\kappa_{ee}$)
of helium as a function of the degeneracy
parameter $T/\TF$: comparison of \citet{HL} data
({\em crosses}) with the present $ei+ee$ results ({\em solid lines})
and with the $ei$ opacities ({\em dashed lines})
for values of $\log T$ (K) marked near the lines.
\label{fig:HL_He1}}
\end{figure}

There is no such procedure for the $ee$ collisions.
Note that a correct expression for $\nu_{ee}$
must reproduce
the general dependence (\ref{nu_ee_nondeg}) at very high $T$ or 
very low $\rho$
(maybe up to a factor of a few, given the logarithmic accuracy
of the estimates of $\Lambda$
in \S\,\ref{sect:nondeg}).
In particular, \req{nu_ee_nondeg} implies
$\nu_{ee} \propto T^{-3/2} \ln T$ at $T\to\infty$,
which agrees with an accurate asymptote at $T\gg\TF$
obtained 
by R.~H. Williams \& H.~E. DeWitt (1968, unpublished)
and reproduced by Eq.~(16) of \citet{Lampe68}.
In contrast, the fit (\ref{ee-fit}), being substituted 
into \req{nu_ee_deg}, produces a formal asymptote
$\nu_{ee} \propto T^{-1} \ln T$ at $T\to\infty$. Clearly,
\req{ee-fit} must be corrected at $T \gtrsim\TF$.

We construct such a correction, using the numerical data 
published by \citet{HL}, in such a way as to reproduce
their tabulated conductivities in those $\rho-T$ domains
where their data can be trusted (see below).
Specifically, we propose the following interpolation 
   formula\footnote{This represents an improvement over the formula used
   recently by \citet{BP06}.} for the case of arbitrary
degeneracy:
\begin{equation}
 \nu_{ee} = \nu_{ee}^{\mathrm{deg}} \frac{1+t^2}{1+t+b t^2\sqrt{T/\TF }},
\label{nu_ee_fit}
\end{equation}
where $t=25 T/\TF $, 
and $\nu_{ee}^{\mathrm{deg}}$ is given 
by Eqs.~(\ref{nu_ee_deg}) and (\ref{ee-fit}).
The coefficient $b=135/\sqrt{32\pi^7}\approx0.434$
corresponds to
the ratio of the exact limiting expressions at $T\ll \TF $
and $T\gg \TF $ (e.g., Eqs.~(16) and (17) of \citealt{Lampe68}). 

\begin{figure} \epsscale{1}
\plotone{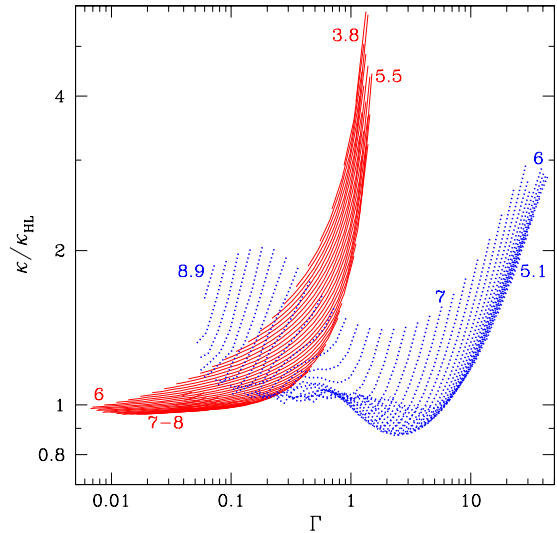}
\caption{Ratio of our new conductive opacities to \citet{HL}
data for helium
as a function of the ion Coulomb coupling parameter $\Gamma$
for $10^{3.8} \leq T({\rm K}) < 10^9$. 
Each line corresponds to a constant $T$ value. 
{\em Solid lines}: nondegenerate matter; {\em dotted lines}: degenerate matter.
Numbers near some of the curves are the corresponding values of $\log T$\,(K).
\label{fig:HL_He_dif}}
\end{figure}

This approximation is compared to
the numerical tables of \citet{HL}
for helium
in Figs.~\ref{fig:HL_He1} and \ref{fig:HL_He_dif}.
The
agreement is good at $T\gg \TF $. 
In Fig.~\ref{fig:HL_He1} we also show 
the opacities calculated without taking $ee$ scattering into account
(the dashed lines).
In this case, there is a noticeable difference at $T>\TF $.

%%% FIG 2

In Fig.~\ref{fig:HL_He_dif} 
the solid and dotted lines correspond to the two parts of
the \citet{HL} tables: for weakly ($T/\TF>1.7$) and strongly 
($T/\TF<1.7$) degenerate
electrons, respectively. 
We have chosen this division because, for many $T$ values,
their tables consist of
two disconnected parts, which merge roughly at 
$T/\TF\gtrsim1.7$.
Figure~\ref{fig:HL_He_dif} shows that
there are two domains of strong disagreement
between our opacities and \citeauthor{HL}'s at $\Gamma\gtrsim1$.
These deviations take place near the $\rho-T$ domains
excluded from the \citeauthor{HL} tables
because either the ions are strongly coupled, or their motion
is quantized. In these cases the \citeauthor{HL} theory breaks down,
therefore this disagreement does not invalidate \req{nu_ee_fit}.
An additional disagreement at $\Gamma\gtrsim1$ and
relatively low temperatures comes from the correction
by \citet{SY06}, discussed above.

\subsection{Summary of Recent Updates of the Conductive Opacities}
\label{sect:sum}
{}From the preceding discussion it is clear that 
the previous conductive opacity evaluations had the following
limitations.

\begin{itemize}
\item
Formulae by \citet{I83,I84,I93} and \citet{PBHY} are relevant only at strong degeneracy, 
$T\ll \TF $. This condition is violated in the outer parts
of RGB cores \citep{Catelan}.

\item
\citet{I83} relied on the ion structure factor calculations at $\Gamma>1$.
The condition $\Gamma>1$ is also violated in some parts of RGB
stars \citep{cdph96,Catelan}.
However, \citet{PBHY} used more accurate evaluations of the ion structure
factor at $\Gamma\geq1$ and an interpolation at $\Gamma<1$,
which allowed them to consider conductivities at arbitrary $\Gamma$ in spite 
of the fact that detailed calculations for $\Gamma \la 1$ have not yet been 
carried out.

In addition, \citet{PBHY} have taken account of
multiphonon scattering and incipient long-range order
around the melting line ($\Gamma=\Gamma_\mathrm{m}$) on the plasma phase diagram \citep{BKPY},
which led to a considerable correction to the opacities
at $\Gamma\sim100-300$, compared to Itoh et al.\ results.

\item
\citet{P99} employed a specific thermal averaging
for the contribution of the $ei$ scattering
to the conductive opacity
(\S\,\ref{sect:partial}), implemented it in subsequent papers
\citep[e.g.,][]{VP01,GYP}, and created an
online database for computing the updated electrical and thermal plasma
conductivities.\footnote{\url {\tt http://www.ioffe.ru/astro/conduct/}\label{foot:conduct}}
This allows one
to compute conductivities determined by 
$ei$ scattering at arbitrary degeneracy.

However, the absence of a similar thermal averaging 
procedure for the $ee$
scattering prevented one from obtaining accurate results
in the case where the $ee$ scattering can be important.
In the conductivity database, two ways were employed previously 
to extend the non-magnetic opacities beyond the domain of $T\ll\TF$:
(A) to ignore $ee$ scattering altogether,
or 
(B) to continue using the theory by \citet{UY80}.
Model A was employed in the table of opacities
and model B in the Fortran program presented
in the database (the latter model was used in 
the analysis by \citealt{Catelan}).

\item
In model B the treatment of the $ee$ scattering
was based on the old theory of \citet{FI76} and \citet{YU80}.
As stressed in \S\,\ref{sect:ee},
the theory has been considerably improved recently
by \citet{SY06}.

\end{itemize}

In the present paper, we have done the following improvements
to the previous treatments of the contribution to
the conductive opacity due to the $ee$ scattering:

\begin{itemize}
\item
In degenerate matter, the old theory 
\citep{UY80} and corresponding analytical formulae \citep{PCY,PBHY}
are replaced
by the new theory and analytical formulae by \citet{SY06}
(see \S\,\ref{sect:ee}).

\item
In partially degenerate and nondegenerate matter,
extrapolation of the $ee$ opacity according to model B above
is replaced by the interpolation formula 
(\ref{nu_ee_fit}), which correctly reproduces
the high-temperature asymptote (\ref{nu_ee_nondeg})
and minimizes differences with respect to numerical tables
in the domain of weak degeneracy
(see \S\,\ref{sect:partial}).
\end{itemize}

Note that the latter interpolation does not concern the $ei$
contribution to the opacity, which is calculated numerically
according to Eqs.~(\ref{thav-kappa}) and (\ref{thav}).

We have updated previously developed Fortran code for thermal
conductivities of electron-ion plasmas in arbitrary magnetic fields
available online (see footnote~\ref{foot:conduct}) 
by including $ee$ scattering in the particular case of
zero magnetic field, taking into account the recent progress in the
treatment of the $ee$ conductive opacity, summarized above.  
We have also updated the
table of thermal conductivities
of fully ionized electron-ion plasmas, 
calculated using this code and available at the same URL. The table
covers densities $\rho$ from $10^{-6}$ to $10^9$ g cm$^{-3}$, 
temperatures $T$ from $10^3$ to
$10^9$~K, and ion charge numbers $Z_i$ from 1 to 60. 
For convenience of potential users, the table is supplemented by
an interpolation routine in Fortran.

%%% FIG 3

\begin{figure}
\plotone{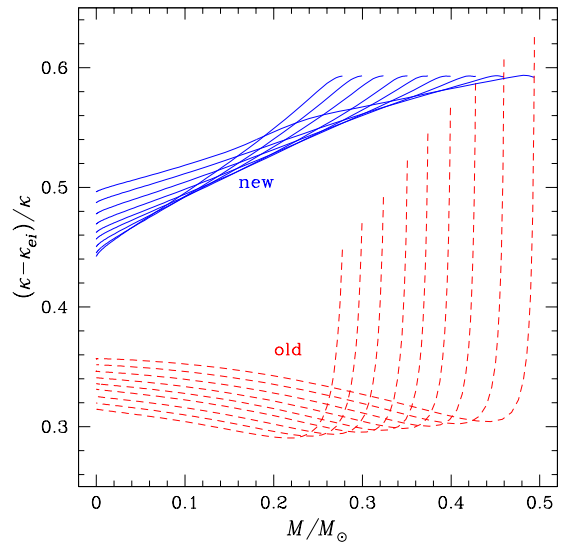}
\caption{Ratio of the $ee$ interaction opacity 
to  the total ($ee+ei$) conductive opacity
along the RGB core models of a $0.8M_{\odot}$ metal poor star 
discussed by \citet{Catelan}, as a function of the Lagrangian 
mass coordinate (in solar units). 
For the $ei$ contribution to the opacity,
the fitting formulae of \citet{PBHY}
are used together with thermal averaging 
according to Eqs.~(\ref{thav-kappa}) and (\ref{thav}). 
For the $ee$ contribution, the {\em solid lines} correspond to 
    Eqs.~(\ref{nu_ee_deg}), (\ref{ee-fit}), and (\ref{nu_ee_fit}),
    and the {\em dashed lines} to the older theory for degenerate
electrons \citep{UY80,PBHY}.
Each curve of a series corresponds to a specific RGB stellar model,
whose He core mass is equal to the value of the mass attained at the end of the curve.
\label{fig:RGB_ee}}
\end{figure}

In Fig.~\ref{fig:RGB_ee} we show the ratio of the 
$ee$ interaction opacity to the total conductive opacity 
in the RGB core models
discussed by \citet{Catelan}. The new results ({\em solid lines}), which 
correspond to \req{nu_ee_fit}, are compared 
with the old model B ({\em dashed lines}, which is similar to 
Fig.~19 in \citealt{Catelan}). 
The differences between the new and old results are caused, first, by
the better treatment of $\nu_{ee}$ in the regime of partial degeneracy
according to \req{nu_ee_fit}, which decreases the opacity
at $T\gg\TF$ but increases it at $T\sim\TF$, 
and secondly, by the improved treatment of $\nu_{ee}^\mathrm{deg}$,
reviewed in \S\,\ref{sect:ee}, which increases the opacity
at high densities. The opacity decrease is noticeable only
in the outer layers of the RGB stellar models with relatively
high $M_\mathrm{tot}$, whereas the increase is
overwhelming in the most of the stellar core.
On the whole, in 
the new calculations the $ee$ interactions play an even more important 
role for the physical conditions prevailing in the interiors of low-mass 
RGB stars than had previously been suspected \citep{cdph96,Catelan}. 

%%% FIG 4

\begin{figure}
\plotone{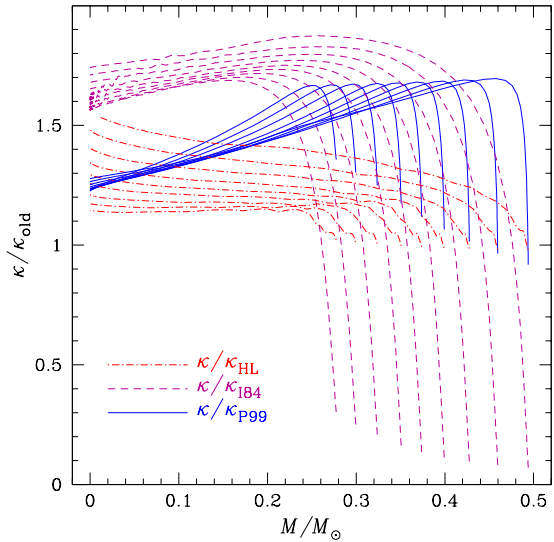}
\caption{Ratio of the $ee+ei$ conductive opacity
$\kappa_\mathrm{c}$ computed in this paper to previous evaluations
by \citeauthor{HL} ($\kappa_\mathrm{HL}$, {\em dot-dashed lines}),
    \citeauthor{I83} ($\kappa_\mathrm{I84}$, {\em dashed lines}),
   and \citeauthor{PBHY} ($\kappa_\mathrm{P99}$, {\em solid lines})
for the same stellar models of Fig.~\ref{fig:RGB_ee}.
\label{fig:RGB_tot}}
\end{figure}

Figure \ref{fig:RGB_tot} shows the ratio of our present conductive
opacities to various approximations
employed previously. One can notice values that are significantly different from unity. 
The difference between our results and \citet{P99} (solid curves) is due to 
the neglect of the $ee$ scattering by the latter.
In the case of the comparison with \citet{HL} (dot-dashed curves) the difference 
is mainly due to the deficiency of their treatment of strongly coupled
and relativistic plasmas, discussed above.
The comparison with \citet{I84} (dashed curves)
shows differences that are partly due to the $ee$ collisions,
but the most striking difference at the end of each curve
(that corresponds to the boundary of the He core) occurs because
$T>\TF$ there, whereas the
\citeauthor{I84}
formulae are valid only for $T\ll\TF$.

%%%%%%%%%%%%%%%%%%%%%%%%%%%%%%%%%%%%%%%%%%%%%%%%%%%%%%%%%%%%%%%%%%%%%%%%
\section{The Effect on the Evolution of low-mass Stars}
\label{sect:effects}
In the following we analyze the 
effect of our new conductive opacities on the evolution of low-mass stars. 
For the tests involving RGB and HB models  
we employed the same stellar 
evolution code and input physics discussed in \citet{pcsc04},
the only difference being the 
use of our new conductive opacities in place of the \citet{P99} ones. 
More in detail, we have computed the evolution of the 
following models: i) $M_\mathrm{tot}=0.8\,M_\odot$, $Z=10^{-4}$, $Y=0.245$, 
ii) $M_\mathrm{tot}=0.8\,M_\odot$, $Z=10^{-3}$, 
$Y=0.246$, iii) $M_\mathrm{tot}=0.85\,M_\odot$, $Z=0.004$, $Y=0.251$, 
and iv) $M_\mathrm{tot}=1.0\,M_\odot$, $Z=0.0198$, $Y=0.2734$. This choice of initial stellar masses and chemical composition
allows us to compare directly these models with the results by \citet{pcsc04}, who employed the \citeauthor{P99} 
opacities.

\subsection{Evolution along the Red Giant Branch}
\label{sect:RGB}
\begin{figure}
\plotone{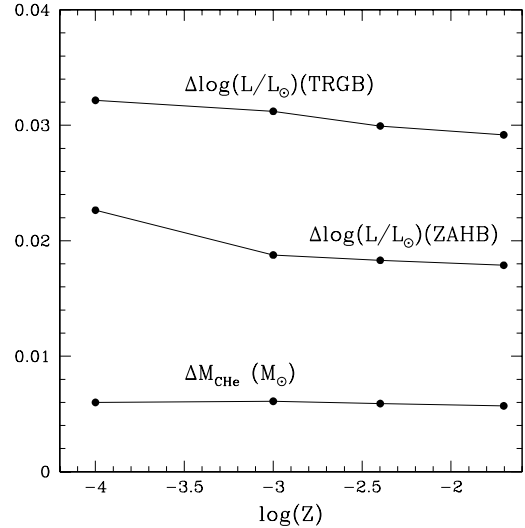} 
\caption{Differences in the He-core mass, RGB tip brightness and ZAHB 
luminosity as a function of the heavy element abundance, between stellar models
computed with the \lq{old}\rq\ conductive opacities by \citet{P99} and our 
new $\kappa_\mathrm{c}$ estimates (see text for details).
\label{fig:difmchel}} 
\end{figure}

The evolution of the models has been followed from the Zero Age Main Sequence until 
He ignition at the RGB tip. We have included 
mass loss according to the \citet{rei75} formula with $\eta=0.4$. 
The new conductive opacities do not affect the
morphology and location of the evolutionary tracks in the H-R diagram: as expected, 
only the He-core mass at the He-flash 
and the corresponding RGB tip 
brightness are modified.
In Table~\ref{rgbtip} 
the main properties of the two sets of RGB models computed by using alternatively the 
\lq{old}\rq\ $\kappa_\mathrm{c}$ predictions and the \lq{new}\rq\ ones are summarized; Fig.~\ref{fig:difmchel} 
displays the differences of the model He-core masses  
and RGB tip brightness as a function of $Z$. 

\begin{deluxetable*}{cccccccccc}
\tablecaption{Selected properties of the models at the RGB tip.}
\tablehead{
\colhead{version} &
\colhead{$t_{\rm TRGB}$\tablenotemark{a}}&
\colhead{$\log(L/L_\odot)$\tablenotemark{b}}&
\colhead{$\log{T_\mathrm{eff}}$\tablenotemark{c}}&
\colhead{$\log{T_c}$\tablenotemark{d}}&
\colhead{$\log{\rho_c}$\tablenotemark{e}}&
\colhead{$M_\mathrm{cHe}$\tablenotemark{f}}&
\colhead{$M(T_\mathrm{max})$\tablenotemark{g}}&
\colhead{$\log{T_\mathrm{max}}$\tablenotemark{h}} &
\colhead{$M_I$\tablenotemark{i}} } 
\startdata
\multicolumn{10}{c}{$0.8\,M_\odot - Z=0.0001 - Y=0.245$}\\ 
\hline
 new & 12.49 & 3.2586 & 3.6427 & 7.863 & 6.061 & 0.4971 & 0.178 & 7.973 & $-4.085$ \\
 old & 12.48 & 3.2908 & 3.6411 & 7.870 & 6.079 & 0.5031 & 0.170 & 7.968 & $-4.164$ \\
\hline
\multicolumn{10}{c}{$0.8\,M_\odot - Z=0.001 - Y=0.246$}\\ 
\hline
 new & 13.69 & 3.3384 & 3.5958 & 7.870 & 6.020 & 0.4852 & 0.162 & 7.969 & $-4.147$ \\
 old & 13.68 & 3.3696 & 3.5946 & 7.877 & 6.039 & 0.4913 & 0.154 & 7.966 & $-4.223$ \\
\hline
\multicolumn{10}{c}{$0.85\,M_\odot - Z=0.004 - Y=0.251$}\\ 
\hline
 new & 13.63 & 3.3829 & 3.5473 & 7.873 & 5.997 & 0.4788 & 0.167 & 7.975 & $-3.933$ \\
 old & 13.62 & 3.4128 & 3.5453 & 7.880 & 6.015 & 0.4847 & 0.161 & 7.970 & $-3.989$ \\
\hline
\multicolumn{10}{c}{$1.0\,M_\odot - Z=0.0198 - Y=0.2738$}\\ 
\hline
 new & 12.52 & 3.4168 & 3.4822 & 7.874 & 5.954 & 0.4669 & 0.191 & 7.978 & $-2.266$ \\
 old & 12.50 & 3.4459 & 3.4766 & 7.881 & 5.972 & 0.4726 & 0.128 & 7.977 & $-2.239$ \\
\hline
\enddata 
\tablenotetext{a}{{}Age (in Gyr) at the RGB tip.}
\tablenotetext{b}{{}Logarithm of the luminosity (in solar units) at the RGB tip.}
\tablenotetext{c}{{}Logarithm of the effective temperature at the RGB tip.}
\tablenotetext{d}{{}Logarithm of the central temperature (in K) at He ignition.}
\tablenotetext{e}{{}Logarithm of the central density (g cm$^{-3}$) at He ignition.}
\tablenotetext{f}{{}He-core mass (in solar units) at He ignition.}
\tablenotetext{g}{{}Location in mass (in solar units) of the off-center temperature maximum at the RGB tip.}
\tablenotetext{h}{{}Logarithm of the maximum off-center temperature (in K) at He ignition.}
\tablenotetext{i}{{}Absolute $I$-Cousins magnitude of the RGB tip.}
\label{rgbtip} 
\end{deluxetable*}

%%% FIG 5

\begin{figure*}
\plotone{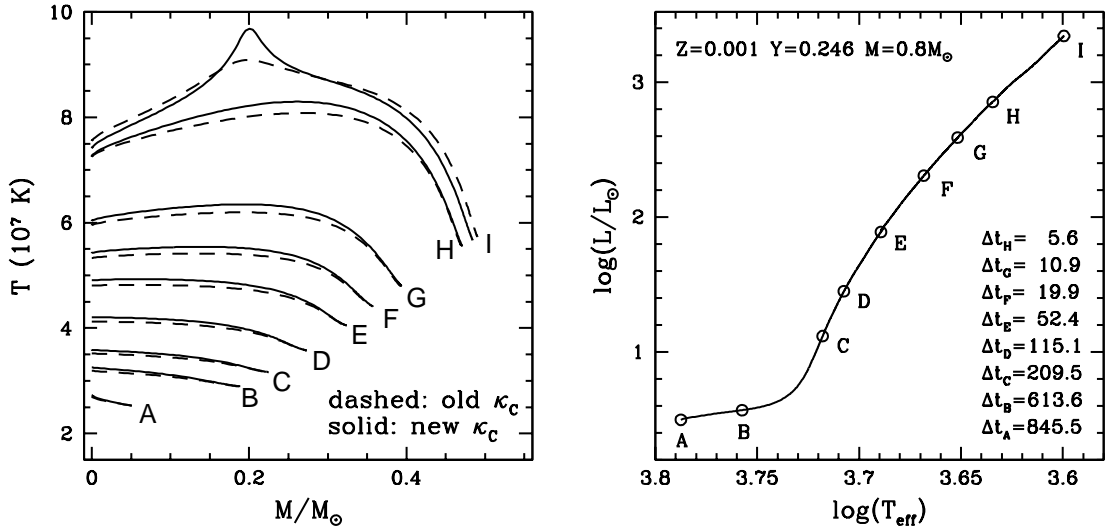}
\caption{\emph{Left panel}:
thermal stratification of the He core at different stages along the RGB 
evolution of a $0.8\,M_\odot$, $Z=0.001$, $Y=0.246$ model, computed with 
the \lq{old}\rq\ conductive opacities by \citet[][{}\emph{dashed lines}]{P99} and our 
new computations ({\em solid lines}) respectively. The hottest sequences   
correspond to ignition of He at the tip of the RGB. 
\emph{Right panel}: location in the H-R diagram of the 
stellar models (labelled with capital letters) whose thermal stratification is shown in the left panel.
For each model, the time (in Myr) needed to reach the He flash at the RGB tip is also listed.
\label{fig:strut}} 
\end{figure*}

Our new $\kappa_\mathrm{c}$ values provide He-core masses at the He-flash 
   lower than in the case when the previous $\kappa_\mathrm{c}$ calculations
   are used (model A described in \S\,\ref{sect:sum}). The difference amounts to
$0.006\,M_\odot$, regardless of the value of $Z$. The reason is
that the new $\kappa_\mathrm{c}$ values are larger than the 
older ones, thus producing a different thermal stratification in the He core. 
The stellar models based on the updated
$\kappa_\mathrm{c}$ at the RGB tip are cooler at the center of the star but hotter at the 
point of the maximum off-center temperature where
the He-ignition occurs. This
is clearly shown by Fig.~\ref{fig:strut} (left panel) that displays the thermal stratification 
along the He core, at 
different stages along the RGB, for a selected metallicity. The right panel in the same
figure shows the correspondence between the various internal
structures and the location of the models along 
the RGB in the H-R diagram.

%%% FIG 6 

The stellar models based on the new
$\kappa_\mathrm{c}$ achieve the thermal conditions required for the He-ignition
at a slightly higher age (see Table~\ref{rgbtip}) compared to the 
model A mentioned above. This is because the new conductive opacities  
force the stellar structures to have a different temperature gradient in
the He core. As a consequence, the temperature at the base of the
H-burning shell is slightly lower than for models based on the old
$\kappa_\mathrm{c}$ determinations (see data shown in Fig.~\ref{fig:strut}).

The difference in He-core mass at the RGB tip causes a change of  
the RGB tip brightness: new models appear to be fainter by
$\Delta \log {(L/L_\odot)}\approx0.03$. This change affects distances 
based on the RGB tip brightness. 
It is well known \citep[see, e.g.,][]{lfm93,sc98} that the $I$-band (Cousins) brightness 
of the RGB tip ($M_{I}^{\rm TRGB}$) is a powerful 
distance indicator for
old, metal-poor and intermediate-metallicity stellar systems such as Local Group Dwarfs, because 
it is  very weakly sensitive to the heavy 
element abundance (and age, for ages above a few Gyr). 
For $\rm [M/H]$ (${\rm [M/H]}=\log(Z/X)-\log(Z/X)_{\odot}$) ranging between 
$-2.0$ and $-0.6$, $M_{I}^{\rm TRGB}$ changes by about 0.15~mag. 
The RGB tip brightness being fixed by the He-core mass at the He-flash, 
any change in $M_\mathrm{cHe}$ translates into a variation in $M_{I}^{\rm TRGB}$.

The data in Table~\ref{rgbtip}
show how the new $\kappa_\mathrm{c}$ decreases the $I$-band brightness 
of the RGB tip by about $0.07-0.08$~mag for metallicities lower than about
$Z\sim0.004$. At solar metallicity, the behavior reverses and the new
estimate is brighter by $\sim0.03$~mag. 
This reversal of the trend is entirely due to the trend of bolometric correction $BC_I$ 
with $T_\mathrm{eff}$, 
which overcompensates the smaller bolometric luminosity at the RGB tip. In fact, 
all stellar tracks based on the new $\kappa_\mathrm{c}$ experience the He-ignition at fainter bolometric
magnitudes and higher effective temperatures. The higher $T_\mathrm{eff}$ values are caused by 
the larger (of the order of 
$0.01\,M_{\odot}$) total mass at the flash, which is a consequence of the smaller amount of 
mass lost along the RGB evolution, due to the shorter RGB evolutionary lifetime 
and lower RGB tip luminosity of the models.
Because of the trend of $BC_I$ with $T_\mathrm{eff}$, 
the final value of $M_{I}^{\rm TRGB}$ gets brighter at this high metallicity. 

The $M_{I}^{\rm TRGB}$ values obtained with our new opacities turn out to be within 
about 1$\sigma$ of the empirical value obtained by \citet*{bellaz}, who coupled their own 
detection of the RGB tip in the Galactic globular cluster $\omega$~Centauri (NGC~5139) to the cluster 
distance obtained by \citet{tomp01} from the analysis of the 
cluster eclipsing binary OGLEGC~17. 

Before closing this section, we wish to note that the main source of
change in the value of $M_\mathrm{cHe}$ when moving from the
$\kappa_\mathrm{c}$
evaluations provided by \citet{P99} to our new conductive opacities is 
related to the inclusion of the $ee$ interactions in our new treatment 
of conductive transport. 
More in detail, proper treatment of the $ee$ interactions
according to the fit (\ref{nu_ee_fit}), but with the old
treatment of the $ee$ scattering frequency in degenerate matter
$\nu_{ee}^{\mathrm{deg}}$ \citep{UY80,PCY,PBHY},
contributes about 70\% (i.e., $\sim0.004\,M_\odot$) of the change in $M_\mathrm{cHe}$. 
The better treatment of $\nu_{ee}^{\mathrm{deg}}$ that takes into
account the relativistic effect of exchange of
transverse plasmons (see \S\ \ref{sect:ee})  
accounts for the remaining one-third of the total
variation.

\subsection{Evolutionary Properties during the Core He-Burning Phase}

As discussed in the Introduction, any change of the He-core mass at 
He ignition causes a variation in the luminosity of the Zero Age
Horizontal Branch (ZAHB). It is worthwhile to investigate how the 
change in $M_\mathrm{cHe}$ brought about by the new opacities affects the 
properties of HB stellar models. To this purpose, we computed a set 
of HB models originated from the RGB progenitors discussed 
above.

We focus our attention on the HB models whose ZAHB location is at $\log(T_\mathrm{eff})=3.85$. We choose this 
   effective temperature value\footnote{We note that this $T_{\rm eff}$ value 
   cannot be considered fully representative of the average temperature in the 
   RR Lyrae strip (see for instance, \citealt{marconi}). 
   However, this has no impact on the present discussion.} for consistency with our previous analysis on this issue. 

The main properties of these models are listed in our Table~\ref{zahb}. 
The ZAHB models with the updated $\kappa_\mathrm{c}$ are
fainter than those based on the previous estimates, 
because of the lower He-core mass of the progenitor. 
The difference is $\Delta{\log(L/L_\odot)}\sim 0.02$ within the explored
metallicity range. The corresponding $V$-band ZAHB brightness 
($M_V({\rm ZAHB})$) of the models is reported in Table~\ref{zahb}. 
The new conductive opacities cause an increase in 
$M_V({\rm ZAHB})$ ranging 
from $\sim0.06$~mag at $Z=0.0001$ down to $\sim 0.04$~mag at solar metallicity. 
This has the effect of slightly decreasing the distance to 
Galactic globular clusters and of systematically 
increasing their ages by  $\sim0.7$~Gyr.
These changes in the globular cluster age and distance scales 
are within current uncertainties on these estimates 
(for more details, see, e.g., \citeauthor{recioblanco} \citeyear{recioblanco}; 
\citeauthor{deangeli} \citeyear{deangeli}, and references therein).

%%% Table 2 %%%

%\clearpage
\begin{deluxetable}{ccccc}
\tablecaption{Selected properties of HB models.}
\tablehead{
\colhead{version} &
\colhead{$M_{3.85}$\tablenotemark{4}}&
\colhead{$\log(L/L_\odot)_{3.85}$\tablenotemark{b}}&
\colhead{$M_V$\tablenotemark{c}}&
\colhead{$t_{\rm HB}$\tablenotemark{d}} } 
\startdata
\multicolumn{5}{c}{RGB progenitor: $0.8\,M_\odot - Z=0.0001 - Y=0.245$}\\ 
\hline
 new & 0.785 & 1.734 & 0.434 & 83.55 \\
 old & 0.801 & 1.757 & 0.377 & 79.03 \\
\hline
\multicolumn{5}{c}{RGB progenitor: $0.8\,M_\odot - Z=0.001 - Y=0.246$}\\ 
\hline
 new & 0.629 & 1.645 & 0.628 & 94.61 \\
 old & 0.639 & 1.664 & 0.582 & 89.64 \\
\hline
\multicolumn{5}{c}{RGB progenitor: $0.85\,M_\odot - Z=0.004 - Y=0.251$}\\ 
\hline
 new & 0.577 & 1.575 & 0.825 & 102.98 \\
 old & 0.585 & 1.593 & 0.750 & 100.69 \\
\hline
\multicolumn{5}{c}{RGB progenitor: $1.0\,M_\odot - Z=0.0198 - Y=0.2738$}\\ 
\hline
 new & 0.531 & 1.445 & 1.066 & 115.88 \\
 old & 0.538 & 1.463 & 1.029 & 111.53 \\
\hline
\enddata 
\tablenotetext{a}{{}Mass (in solar units) of the ZAHB model located at $\log{T_\mathrm{eff}}=3.85$.} 
\tablenotetext{b}{{}Logarithm of the luminosity (in solar units) of the ZAHB model located at $\log{T_\mathrm{eff}}=3.85$.}
\tablenotetext{c}{{}Absolute visual magnitude of the ZAHB model located at $\log{T_\mathrm{eff}}=3.85$.}
\tablenotetext{d}{{}Core He-burning lifetime (in Myr) of the HB structure with ZAHB location at $\log{T_\mathrm{eff}}=3.85$.}
\label{zahb} 
\end{deluxetable}

As to the location in effective temperature of the ZAHB models, we have 
    already shown that for a given RGB progenitor and for fixed mass-loss 
    efficiency along the RGB, the total stellar mass at He ignition increases. 
This occurrence -- combined with the decrease of the He core mass -- will 
cause the ZAHB model location
to be slightly cooler with respect to the model based on the old 
opacity predictions.

A consequence of the lower HB luminosity is that the
evolutionary lifetime of the core He-burning phase is slightly longer, by 
$\sim$ 5\,--\,6\%. Together with the lower luminosity of the 
RGB tip and the fainter ZAHB, this might affect the 
theoretical calibration of the $R$-parameter, i.e., the ratio of the 
number of HB stars to the number of 
RGB objects brighter than the ZAHB level \citep[][]{iben68} that is often 
used as an indicator of the initial He abundance 
of Galactic globular clusters. 
Recently, \citet*{csi03} and \citet{salaris04} have derived a new calibration of $R$ 
(using the \citealt{P99} conductive opacities)
that, applied to a large and homogeneous 
database of empirical values 
for Galactic globulars, provides an initial  He-abundance in very good agreement 
with the determination based on recent analysis of 
Cosmic Microwave Background fluctuations and Big Bang nucleosynthesis predictions
(see \citealt{spergel2} and references therein). 
Here we have estimated the value of the $R$ parameter based on the models 
presented in the previous sections and 
compared the results with ones based on the \citeauthor{P99} opacities. 
    The use of the new $\kappa_\mathrm{c}$ increases the HB lifetime ($t_{\rm HB}$) 
    by $\sim 5\%$, but the time spent by the RGB progenitor at luminosities 
    higher than the ZAHB level also increases by $\sim3.5\%$.
The net effect is that, for a fixed initial chemical composition, the value of $R$ changes 
by just $+0.03$ ($\sim2$\%).
The dependence of the $R$-parameter on the He content is $\delta{R}/\delta{Y}\sim 10$ 
\citep[see][]{csi03} and  
therefore the change of the $R$ calibration 
caused by the new $\kappa_\mathrm{c}$ would lead to a variation by only 0.003 of the estimated 
initial He content of the Galactic globular cluster system. This means that 
the reliability of the $R$-parameter 
calibration presented by \citeauthor{csi03} and \citeauthor{salaris04} 
    is not seriously affected 
by the revised conductive opacities.

\subsection{Very Low-Mass Stars and White Dwarfs}

In very low-mass (VLM) stars, i.e. structures with mass lower than $\sim 0.3-0.4\,M_\odot$, 
the temperature is of the order of the electron Fermi temperature. 
More in detail, considering  
$T/T_{F}\approx 3.31\times10^{-6} T (\mu_e/\rho)^{2/3}$ together with values of temperature 
and density characteristic of the interiors
of VLM stars, one obtains $T/T_F\sim1-2$. This means that the VLM stellar interiors 
are under conditions of partial electron degeneracy, 
the degeneracy level strongly decreasing with the total mass.
Taking into account such circumstantial evidence, we decided to test the impact of the 
new opacity estimates on the structure and evolution  
of VLM stars. We adopted the same evolutionary code and physical inputs 
as in \citet{c3pz} but including our new 
conductive opacities. We computed VLM stellar models in the  mass range 
$0.09 - 0.15\,M_\odot$ for a metallicity $Z=0.001$. 
A close comparison with the corresponding evolutionary models provided by \citet{c3pz}, based on the
\citet{I93} conductive opacities, shows that -- at least in the considered range of stellar mass --
there are negligible differences between the two sets of 
stellar models concerning both the structural properties and evolutionary lifetimes. 

The extremely degenerate cores of white dwarfs are properly covered by 
the Itoh et al.\ and Yakovlev et al.\ opacities, whereas the \citet{HL} opacities are 
not adequate in this 
regime \citep[see, e.g.,][]{PMS02}. 
However, it is well known \citep[see, e.g.,][]{han99, sal00, PMS02} that neither the 
Itoh et al.\ nor the \citeauthor{HL} conductive opacities can completely cover the 
He-envelopes of white dwarfs during their entire cooling phase. A solution generally used  
\citep[see, e.g., the models by ][]{han99, sal00} is to employ the 
Itoh et al.\ (or Yakovlev et al.)
opacities, supplemented outside their range of validity (e.g., when the 
matter is only weakly degenerate) by the \citeauthor{HL} data. 
Given that  the different opacity sources are not based on the same assumptions and
input physics 
(see \S\S\,\ref{sect:intro}, \ref{sect:cond}) 
their matching in the regime of weak degeneracy is sometimes
problematic.

It should be stressed that the \citet{PBHY} opacities,
calculated a few years ago 
and aimed especially at white dwarf and neutron star
applications,
have already superseded the older results of Itoh et al., as explained in
\S\,\ref{sect:cond}. Nevertheless, even the \citeauthor{PBHY} opacities 
should be corrected following the improvement of the treatment
of the $ee$ scattering contribution described in \S\S\,\ref{sect:ee}
and \ref{sect:partial}.

Here we test the impact of the new opacities
described in \S\,\ref{sect:cond} and available at the URL 
referenced in footnote~\ref{foot:conduct} on the cooling timescale of white dwarfs. 
We emphasize that the new calculations adequately cover the whole white 
dwarf structure during its cooling sequence. We employed the 
same code and input physics 
as in \citet{sal00} to compute the cooling evolution of a 
$0.54\,M_{\odot}$ and a $1.0\,M_{\odot}$ DA 
white dwarf with a pure carbon core 
\citep[the thickness of the He and H envelope layers is as in][]{sal00} using both our 
new opacities and the combination of 
Itoh et al.\ and  \citet{HL} opacities mentioned before. 
The mass-radius relationship is unchanged, and
the cooling times at a given luminosity 
turn out to be almost exactly the same, irrespective of the opacities used. The maximum 
differences are of the order of 
just 1\% (the cooling times obtained with the new opacities being longer) when 
$\log(L/L_\odot) < -$5.0 and the cooling ages 
are above 14~Gyr.

%%%%%%%%%%%%%%%%%%%%%%%%%%%%%%%%%%%%%%%%%%%%%%%%%%%%%%%%%%%%%%%%%%%%%%%%
\section{Summary}
\label{sect:concl}

We have reviewed the basic theory to estimate the electron-conduction 
opacity for the stellar matter, and discussed limitations and shortcomings of the
existing calculations that are employed in stellar evolution modelling. 
We present new results that can be applied to both partial (e.g., in the cores of RGB stars, envelopes 
    of white dwarfs) and high degeneracy (e.g., white dwarf cores) regimes, for an arbitrary chemical composition. 
Our results update the 
previous calculations by \citet{HL}, Itoh and coworkers \citep[e.g.,][]{I83,I84,I93},
and Yakovlev and coworkers \citep[e.g.,][]{PBHY,P99,SY06},
extensively used in the literature. They improve upon \citet{HL} by including an updated 
treatment of both the $ei$ scattering and the $ee$ scattering for strongly coupled 
and relativistic plasmas, and are not restricted to the specific mixtures published 
by \citeauthor{HL}. 
Differences with Itoh and coworkers are mainly due to their neglect of the $ee$ scattering,
inaccurate treatment of the $ei$ scattering at $\Gamma\sim100-200$,
and the fact that their results do not extend to the regimes of partial degeneracy
($T>\TF$) and weak ion coupling ($\Gamma<1$).
Differences with \citet{P99} are essentially due to his neglect of the $ee$ scattering.
Compared to the latest results of Yakovlev's group
\citep{PBHY,SY06}, our new opacities differ essentially 
in the extension of the $ee$ scattering results to the regime of partial degeneracy.
We compared our new computations with the aforementioned sets of opacities in the temperature-density 
regime of RGB star cores, and found non-negligible differences. On the whole, the \citet{HL} results 
are the most similar to ours in the RGB cores, the maximum differences being equal to $\sim$ 50\% in the 
central regions of the cores. 

The effect of using our new opacities for RGB and HB calculations instead of the 
\citet{P99} data is to decrease the bolometric luminosity of the RGB tip and the ZAHB by 
$\Delta \log(L/L_{\odot}) \sim 0.03$ and $\Delta \log(L/L_{\odot}) \sim 0.02$, respectively. The He-core masses at 
the RGB tip are decreased by $\sim$ 0.006 $M_{\odot}$, the total 
RGB evolutionary timescales are almost unchanged, 
but the time spent along the RGB above the HB increases by $\sim$3 \%, and the HB evolution 
is longer by $\sim$ 5\%. The calibration of the $R$-parameter as a function of $Y$ remains almost unchanged. 

When our new opacities are applied to white dwarf models, the resulting mass-radius relationship 
and cooling times are essentially 
unchanged compared to the results obtained employing the combination of Itoh and \citet{HL} opacities 
that until now was needed to cover the whole white dwarf structures.
The impact on Main Sequence models of very-low-mass stars is also negligible.

\acknowledgments
A.P.\ is grateful to Dima Yakovlev, Andreas Reisenegger, Peter Shternin,
and Stephanie Hansen for useful discussions and remarks. 
The work of A.P.\ is supported in part by FASI (Rosnauka) grant 
NSh-9879.2006.2, by RFBR grants 05-02-16245 and 05-02-22003, and by the 
Visiting Professors Program of Pontificia Universidad Cat\'olica de Chile. 
M.C.\ acknowledges support by Proyecto FONDECYT Regular No.~1030954.


\begin{thebibliography}{99}

\bibitem[Baiko \& Yakovlev(1995)]{BY95}
Baiko, D.~A., \& Yakovlev, D.~G. 1995,
Astron.\ Lett., 21, 702

\bibitem[Baiko et al.(1998)]{BKPY}
Baiko, D.~A., Kaminker, A.~D., Potekhin, A.~Y., \& Yakovlev, D.~G. 1998,
\prl, 81, 5556

\bibitem[Barriga-Carrasco \& Potekhin(2006)]{BP06}
Barriga-Carrasco, M.~D., \& Potekhin, A.~Y. 2006,
Laser and Particle Beams, 24, 553

\bibitem[Bellazzini et al.(2001)Bellazzini, Ferraro, \& Pancino]{bellaz} 
Bellazzini, M., Ferraro, F.~R., \& Pancino, E., 2001, 
\apj, 556, 635 

\bibitem[Braginski{\u\i}(1958)]{Brag}
Braginski{\u\i} S.~I. 1958,
% Zh.\ Eksp.\ Teor. Fiz.1957 {33}, 459
% (Engl.\ transl.: 
Sov.\ Phys. JETP, {6}, 358%)

\bibitem[Brown et al.(2002)Brown, Bildsten, \& Chang]{BBC}
Brown, E.~F., Bildsten, L., \& Chang, P. 2002,
\apj, 574, 920

\bibitem[Carson(1976)]{Carson}
Carson, T. R. 1976,
% Stellar opacity
\araa, {14}, 95%--117

\bibitem[Cassisi et al.(2000)]{c3pz}
Cassisi, S., Castellani, V., Ciarcelluti, P., Piotto, G., \& Zoccali, M. 2000, \mnras, 315, 679

\bibitem[Cassisi et al.(2003)Cassisi, Salaris, \& Irwin]{csi03} 
Cassisi, S., Salaris, M., \& Irwin, A.~W. 2003, 
\apj, 588, 852 

\bibitem[Castellani \& Degl'Innocenti (1999)]{cd99} Castellani, V., \& Degl'Innocenti, S. 1999, \aap, 344, 97

\bibitem[Catelan(2005)]{Catelan}
Catelan, M. 2005, in Resolved Stellar Populations, ed. D. Valls-Gabaud \& M. Ch\'avez 
(San Francisco: ASP), in press (astro-ph/0507464) 

\bibitem[Catelan et al.(1996)Catelan, de Freitas Pacheco, \& Horvath]{cdph96} 
Catelan, M., de Freitas Pacheco, J.~A., \& Horvath. J.~E. 1996, 
\apj, 461, 231

\bibitem[Chabrier \& Baraffe(2000)]{CB00}
Chabrier, G., \& Baraffe, I. 2000,
\araa, 38, 337

\bibitem[De Angeli et al.(2005)]{deangeli}
De Angeli, F., Piotto, G., Cassisi, S., Busso, G., Recio-Blanco, A.,
Salaris, M., Aparicio, A., \& Rosenberg, A. 2005, AJ, 130, 116

\bibitem[Flowers \& Itoh(1976)]{FI76}
Flowers E., \& Itoh, N. 1976,
\apj, {206}, 218

\bibitem[Flowers \& Itoh(1979)]{FI79}
Flowers E., \& Itoh, N. 1979,
\apj, {230}, 847

\bibitem[Flowers \& Itoh(1981)]{FI81}
Flowers E., \& Itoh, N. 1981,
\apj, {250}, 750

\bibitem[Gnedin et al.(2001)Gnedin, Yakovlev, \& Potekhin]{GYP}
Gnedin, O.~Y., Yakovlev, D.~G., \& Potekhin, A.~Y. 2001,
MNRAS, 324, 725

\bibitem[Hansen(1999)]{han99}
Hansen, B.~M.~S. 1999, \apj, 520, 680

\bibitem[Heiselberg \& Pethick(1993)]{HeiselbergPethick}
Heiselberg, H., \& Pethick, C.~J. 1993,
\prd, 48, 2916

\bibitem[Hubbard(1966)]{Hubbard66}
Hubbard, W.~B. 1966,
\apj, 146, 858

\bibitem[Hubbard \& Lampe(1969)]{HL}
Hubbard, W.~B., \& Lampe, M. 1969,
\apjs, 18, 297

\bibitem[Iben(1968)]{iben68}
Iben, I. Jr. 1968, \nat, 220, 143

\bibitem[Itoh \& Kohyama(1993)]{IK93}
Itoh, N., \& Kohyama, Y. 1993,
\apj, 404, 268; erratum: 1994, \apj, 420, 943

\bibitem[Itoh et al.(1983)]{I83} 
Itoh, N., Mitake, S., Iyetomi, H., \& Ichimaru, S 1983, \apj, 273, 774 

\bibitem[Itoh et al.(1984)]{I84}
Itoh, N., Kohyama, Y., Matsumoto, N., \& Seki, M. 1984,
\apj, {285}, 758

\bibitem[Itoh et al.(1993)Itoh, Hayashi, \& Kohyama]{I93}
Itoh, N., Hayashi, H., \& Kohyama, Y. 1993,
\apj, {418}, 405; erratum: 1994, \apj, 436, 418

\bibitem[Lampe(1968)]{Lampe68}
Lampe, M. 1968, Phys.\ Rev., 174, 276

\bibitem[Lee et al.(1993)Lee, Freedman, \& Madore]{lfm93}
Lee, M.G., Freedman, W., \& Madore, B.F., 1993, \apj, 417, 553 

\bibitem[Marconi et al.(2003)]{marconi} Marconi, M., Caputo, F., Di Criscienzo, M., \&
Castellani, M. 2003, \apj, 596, 299

\bibitem[Marshak(1940)]{rm40}
Marshak, R. E. 1940, \apj, 92, 321

\bibitem[Mitake, Ichimaru, \& Itoh(1984)]{mea84}
Mitake, S., Ichimaru, S., \& Itoh, N. 1984, \apj, 277, 375

\bibitem[Pietrinferni et al.(2004)]{pcsc04} 
Pietrinferni, A., Cassisi, S., Salaris, M. \& Castelli, F. 2004, 
\apj, 612, 168

\bibitem[Potekhin(1999)]{P99}
Potekhin, A.~Y. 1999,
\aap, 351, 787

\bibitem[Potekhin \& Chabrier(2000)]{PC00}
Potekhin, A.~Y., \& Chabrier, G. 2000,
\pre, 62, 8554

\bibitem[Potekhin \& Yakovlev(1996)]{PY96}
Potekhin, A.~Y., \& Yakovlev, D.~G. 1996,
\aap, 314, 341

\bibitem[Potekhin et al.(1997)Potekhin, Chabrier, \& Yakovlev]{PCY}
Potekhin, A.~Y., Chabrier, G., \& Yakovlev, D.~G. 1997,
\aap, 323, 415

\bibitem[Potekhin et al.(1999)]{PBHY}
Potekhin A.~Y., Baiko D.~A., Haensel P., \& Yakovlev D.~G. 1999,
\aap, 346, 345

\bibitem[Prada Moroni \& Straniero(2002)]{PMS02}
Prada Moroni, P., \& Straniero, O. 2002,
\apj, 581, 585

\bibitem[Raikh \& Yakovlev(1982)]{RY82}
Raikh, M.~E., \& Yakovlev, D.~G. 1982,
\apss, 87, 193

\bibitem[Recio-Blanco et al.(2005)]{recioblanco} 
Recio-Blanco, A., Piotto, G., de Angeli, F., Cassisi, S., Riello, M., 
Salaris, M., Pietrinferni, A., Zoccali, M., Aparicio, A. 2005, 
\aap, 432, 851

\bibitem[Reimers(1975)]{rei75}
Reimers, D. 1975, Mem. Soc. R. Sci. Liege, 8, 369

\bibitem[Renzini \& Fusi Pecci(1988)]{rfp} 
Renzini, A., \& Fusi Pecci, F. 1988, 
\araa, 26, 199

\bibitem[Salaris \& Cassisi(1998)]{sc98} 
Salaris, M., \& Cassisi, S., 1998, 
\mnras, 298, 166 

\bibitem[Salaris et al.(2000)]{sal00}
Salaris, M., Garc\'ia-Berro, E., Hernanz, M., 
Isern, J., \& Saumon, D. 2000, \apj, 544, 1036

\bibitem[Salaris et al.(2004)]{salaris04} 
Salaris, M., Riello, M., Cassisi, S., \& Piotto, G. 2004, 
\aap, 420, 911

\bibitem[Salaris et al.(2002)Salaris, Cassisi, \& Weiss]{scw02} 
Salaris, M., Cassisi, S., \& Weiss, A. 2002, 
PASP, 114, 375

\bibitem[Shternin \& Yakovlev(2006)]{SY06}
Shternin, P.~S., \& Yakovlev, D.~G. 2006,
\prd, 74, 043004

\bibitem[Spergel et al.(2006)]{spergel2} 
Spergel, D.~N., Bean, R., Dor\'e, O., et al. 2006, \apj, submitted 
(astro-ph/0603449)

\bibitem[Spitzer(1962)]{Spitzer}
Spitzer L., Jr. 1962, Physics of Fully Ionized Gases,
2nd revised edition (Wiley: New York)

\bibitem[Spitzer \& H\"arm(1953)]{SpitzerHarm}
Spitzer, L., Jr., \& H\"arm, R. 1953,
% Transport phenomena in a completely ionized gas
Phys.\ Rev., {89}, 977%--981

\bibitem[Thompson et al.(2001)]{tomp01}
Thompson, I. B., Ka{\l}u{\.z}ny, J., Pych, W., Burley, G., Krzemínski, W., Paczynski, B., 
Persson, S.~E., \& Preston, G. W. 2001, AJ, 121, 3089

\bibitem[Urpin \& Yakovlev(1980)]{UY80}
Urpin, V.~A., \& Yakovlev D.~G. 1980,
\sovast, 24, 126

\bibitem[Ventura \& Potekhin(2001)]{VP01}
Ventura, J., \& Potekhin, A.~Y. 2001,
in {The Neutron Star -- Black Hole Connection}, 
ed.\ C.~Kouveliotou, J.~van Paradijs, \& J.~Ventura, 
NATO ASI Ser.~C, vol.~567 (Dordrecht: Kluwer), 393

\bibitem[Yakovlev(1987)]{Yak87}
Yakovlev, D.~G. 1987,
\sovast, 31, 347

\bibitem[Yakovlev \& Urpin(1980)]{YU80}
Yakovlev, D.~G., \& Urpin, V.~A. 1980,
\sovast, {24}, 303

\bibitem[Young et al.(1991)Young, Corey, \& DeWitt]{YCDW91}
Young, D.~A., Corey, E.~M., \& DeWitt, H.~E. 1991,
\pra, 44, 6508

\bibitem[Ziman(1960)]{Ziman}
Ziman, J.~M. 1960, 
{Electrons and Phonons} 
(Oxford: Oxford University Press)

\end{thebibliography}
\end{document}